\documentstyle[twocolumn,aps]{revtex}

\newcommand {\ch} {\v{c}}

\newcommand{\beq}{\begin{equation}}
\newcommand{\eeq}{\end{equation}}
\newcommand{\bdm}{\begin{displaymath}}
\newcommand{\edm}{\end{displaymath}}
\newcommand{\beqa}{\begin{eqnarray}}
\newcommand{\eeqa}{\end{eqnarray}}
\newcommand{\beqab}{\begin{eqnarray*}}
\newcommand{\eeqab}{\end{eqnarray*}}
\def\nn{\nonumber}
\begin{document}
\draft
\title{Closed $SU(2)_{q}$ invariant spin chain and it's operator content}
\author{S. Pallua\footnote{Electronic address: pallua@phy.hr} and 
P. Prester\footnote{Electronic address: pprester@phy.hr}}
\address{Department of Theoretical Physics, University of Zagreb, POB 162,
Bijeni\ch ka c.32, 10001 Zagreb, Croatia}
\date{\today}
\maketitle
\begin{abstract}
We derive the operator content of the closed $SU(2)_{q}$ invariant quantum
chain for generic values of the deformation parameter $q$.
\end{abstract}

\narrowtext

\subsection{Introduction}
\setcounter{equation}{0}

Integrable models and between them in particular spin chains are very
interesting to study various theoretical ideas. In particular in continuum
limit they describe various relativistic field theories, e.g.~XYZ spin chain
is connected to massive Thirring model \cite{luther}. Possible relevance of
spin chains for high energy QCD was suggested in \cite{faddeev}. Recently
their connection to matrix models was emphasized \cite{lee}.

Between spin chains a particularly interesting class are the quantum group
invariant spin chains. The closed $SU(2)_{q}$ invariant
Hamiltonian was constructed by \cite{martin}. Further investigation showed
that this Hamiltonian implied boundary conditions which depend on the
coupling constant and quantum numbers of the sector \cite{grosse,zapletal}.
This second property together with the property that conformal anomaly was
found to be smaller then one made this Hamiltonian different from the XXZ
chain with the toroidal boundary conditions \cite{alcaraz,grimm}.
The model showed also interesting properties of level crossing including
the change in properties of ground state \cite{grosse}.The properties of
ground state have been investigated by Bethe Ansatz methods. Bethe 
constraint equations have also been written for other groups 
\cite{zapletal,angie1,angie2}. The excited spectrum for $SU(2)_{q}$ was 
previously studied
for rational values of deformation parameter $q$ \cite{pallua} and a
particular class of statistical models belonging to unitary minimal series
was projected. Here we want to address the question of operator content
for generic values of parameter $q$. Here differenly from the rational case
there are no null-states corresponding to ``bad'' representations of
$SU(2)_{q}$ and we shall be able to exploit that representation theory is
isomorphic to the ordinary $SU(2)$.

\subsection{Statistical systems and the quantum chain}

We start with the Hamiltonian for the closed $SU(2)_{q}$ invariant chain 
\cite{martin,grosse}
\beqa
H&=&Lq-\sum_{i=1}^{L-1} R_{i}-R_{0} \label{qham} \\
R_{0} &=& G R_{L-1} G^{-1} \nn \\
G &=& R_{1}\cdots R_{L-1} \nn
\eeqa
where $R_{i}$ are $4\times 4$ matrices
\beqab
R_{i}=\sigma_{i}^{+}\sigma_{i+1}^{-}+\sigma_{i}^{-}\sigma_{i+1}^{+}+
\frac{q+q^{-1}}{4} (\sigma_{i}^{3}\sigma_{i+1}^{3}+1) \\
-\frac{q-q^{-1}}{4} (\sigma_{i}^{3}-\sigma_{i+1}^{3}-2)\; .
\eeqab
We choose the quantum group parameter $q$ to be on the unit circle
\beqab
q&=&e^{i\varphi} \\ \frac{q+q^{-1}}{2}&=&\cos \varphi=-\cos \gamma\; . 
\eeqab
The operator $G$ plays the role of the translation operator
\bdm
G R_{i} G^{-1}=R_{i+1}\; , \quad R_{L}=R_{0}\; ,
\quad i=1,\ldots ,L-1
\edm
and also commutes with the Hamiltonian. Contrary to \cite{pallua}, here we 
shall be interested in the generic case:
\bdm
q^{n}\neq \pm 1 \qquad \forall n\in {\bf Z}
\edm

In this case, one
can decompose the space of states into the direct sum of irreducible
representations of the quantum group which are in one-to-one correspondence
with the usual $SU(2)$ representations. It is therefore sufficient
to treat the highest weight states. All other states can be obtained
with the action of the $S^{-}$ operator. We derived the Bethe
Ansatz (BA) equation in \cite{grosse}. The energy 
eigenvalues are given by
\bdm
E=2\sum_{i=1}^{M}(\cos\varphi -\cos k_{i})\; ,\qquad M=\frac{L}{2}-Q \; .
\edm
Here $Q$ is the eigenvalue of $S^{3}$ and $k_{i}$ satisfy the BA constraints 
\beq \label{qba}
Lk_{i}=2\pi I_{i}+2\varphi (Q+1)-\sum_{\stackrel{\scriptstyle j=1}{j\neq 
i}}^{M}\Theta (k_{i},k_{j})\; ,\quad k_{i}\neq \varphi
\eeq
where $I_{i}$ are integers (half-integers) if $M$ is odd (even), and $\Theta 
(k_{i},k_{j})$ is the usual two-particle phase defined in \cite{grosse}.

Owing to the antisymmetry of phase shifts, from (\ref{qba}) it follows that
\bdm
\sum_{i=1}^{M} k_{i}=\frac{2\pi}{L}\sum_{i=1}^{M} I_{i}+\frac{2M}{L}
\varphi(Q+1)\; .
\edm
This allows us to determine the eigenvalues of the translation operator
$G$ or equivalently of the operator $P$
\bdm
P=i\ln G \; .
\edm
In fact,
\beqab
P&=&\sum_{i=1}^{M} k_{i}-\varphi\left(Q-1+\frac{L}{2}\right) \\&=&
\frac{2\pi}{L}\sum_{i=1}^{M}I_{i}+\varphi\left[-\frac{L}{2}-Q+1+\frac{2M}{L}
(Q+1)\right] \; .
\eeqab

It was also shown in \cite{grosse} that the finite-size correction to the 
thermodynamic limit of the ground-state energy was given by ($L$ even)
\bdm
E_{0}(L)=E_{0}(\infty)-\frac{\pi c\,\zeta}{6L}+O\left(\frac{1}{L}\right)
\edm
where
\bdm
\zeta=\frac{\pi\sin\gamma}{\gamma}\; .
\edm
The conformal anomaly $c$ was found to be
\bdm
c=1-\frac{6(\pi-\varphi)^{2}}{\pi\varphi}
\edm
for $\varphi\in [\frac{\pi}{2},\pi]$ If we parametrize $\varphi$ as:
\beq \label{mvp}
\varphi=\frac{\pi m}{m+1} \; , \qquad m\geq 1\; .
\eeq
then we have ($m$ is irrational)
\beq \label{ccm}
c=1-\frac{6}{m(m+1)}
\eeq
Now we define scaled gaps
\beqab
\bar{E}_{n}&=&\frac{L}{2\pi\zeta}(E_{n}-E_{0}) \\
\bar{P}_{n}&=&\frac{L}{2\pi}(P_{n}-P_{0}+\varphi\, Q)\; .
\eeqab
We introduce the partition function in some sector $Q\geq 0$ :
\bdm
{\cal F}_{Q}(z,\bar{z},L)=\sum_{\rm all\;states}z^{\frac{1}{2}
(\bar{E}_{n}+\bar{P}_{n})}\,\bar{z}^{\frac{1}{2}
(\bar{E}_{n}-\bar{P}_{n})}
\edm
As mentioned above, we are interested in partition function for the highest
weight states:
\beqab
{\cal D}_{Q}(z,\bar{z},L)&=&\sum_{\stackrel{\scriptstyle \rm highest}{\rm
weight\;states}}z^{\frac{1}{2}(\bar{E}_{n}+\bar{P}_{n})}\,
\bar{z}^{\frac{1}{2}(\bar{E}_{n}-\bar{P}_{n})} \\
&=&{\cal F}_{Q}(z,\bar{z},L)-{\cal F}_{Q+1}(z,\bar{z},L)
\eeqab

\subsection{Quantum chain and the XXZ chain with a toroidal 
boundary condition}

To determine ${\cal D}_{Q}$ in the limit $L\longrightarrow\infty$, we can use
results for XXZ chain with a toroidal boundary condition \cite{alcaraz,grimm}.
The Hamiltonian is defined by
\beq \label{tham}
H(q,\phi,L)=-\sum_{i=1}^{L}\left\{ \sigma_{i}^{+}\sigma_{i+1}^{-}+
\sigma_{i}^{-}\sigma_{i+1}^{+}+
\frac{q+q^{-1}}{4} \sigma_{i}^{3}\sigma_{i+1}^{3}\right\}
\eeq
\bdm
\frac{q+q^{-1}}{2}=\cos\varphi =-\cos\gamma
\edm
and
\bdm
\sigma_{L+1}^{\pm}=e^{\mp i\phi}\sigma_{1}^{\pm}\qquad
\phi\in (-\pi,\pi]\; .
\edm
This Hamiltonian commutes with
\bdm
S^{z}=\sum_{i=1}^{L}\sigma_{i}^{3}
\edm
and with the translation operator
\bdm
T=e^{-i\phi\sigma_{1}^{3}/2}P_{1}P_{2}\cdots P_{L-1}
\edm
where $P_{i}$, $i=1,\ldots ,L-1$ are permutation operators
\bdm
P_{i}=\sigma_{i}^{+}\sigma_{i+1}^{-}+\sigma_{i}^{-}\sigma_{i+1}^{+}+
\frac{1}{2}\left( \sigma_{i}^{3}\sigma_{i+1}^{3}+1\right)\; .
\edm
The momentum operator is then
\bdm
P=i\ln T \; .
\edm
The BA constraints for this system are \cite{barber} 
\beq \label{tba}
Lk_{i}=2\pi I_{i}+\phi -\sum_{\stackrel{\scriptstyle j=1}{j\neq
i}}^{M}\Theta (k_{i},k_{j})\; ,\quad i=1,\ldots,M 
\eeq
and give
\beqab
E&=&-\frac{L}{2}\cos\varphi+2\sum_{i=1}^{M}(\cos\varphi-\cos k_{i}) \\
P&=&\sum_{i=1}^{M}k_{i}=\frac{2\pi}{L}\sum_{i=1}^{M}I_{i}+\frac{M}{L}\phi\; .
\eeqab
We define
\beq \label{pl}
\phi=2\pi l\;,\qquad -\frac{1}{2}<l\leq \frac{1}{2} \; .
\eeq
and
\bdm
h=\frac{\pi}{4\varphi}
\edm
Let us denote by $E_{Q;j}^{l}(L)$ and $P_{Q;j}^{l}(L)$ the eigenvalues
of $H$ and $P$ in the sector $S^{z}=Q$ with a boundary condition $l$.
It was shown \cite{grimm} that it was possible to project theories with $c<1$ by
choosing a new ground state with energy $E_{0;j_{0}}^{l_{0}}(L)$. The
number $j_{0}\geq1$ was chosen in such a way that the new ground state gave
the contribution $(z\bar{z})^{h(l_{0}+\nu_{0})^{2}}=(z\bar{z})^{(1-c)/24}$ in
the partition function. So the quantity $(l_{0}+\nu_{0})$ is related to $h$ by
the condition
\bdm
c=1-24h(l_{0}+\nu_{0})^{2}
\edm
where $-\frac{1}{2}<l_{0}\leq\frac{1}{2}$ and $\nu_{0}\in{\bf Z}$. From
(\ref{ccm}) it follows that
\bdm
l_{0}+\nu_{0}=[4hm(m+1)]^{-\frac{1}{2}} \; .
\edm
Now new scaled gaps can be defined as
\begin{mathletters}
\label{fsg} \beqa
\bar{F}_{Q;j}^{k}(L)&=&\frac{L}{2\pi}\left( E_{Q;j}^{k(l_{0}+\nu_{0})}
(L)-E_{0;j_{0}}^{l_{0}}(L)\right) \\ \bar{P}_{Q;j}^{k}(L)&=&
\frac{L}{2\pi}P_{Q;j}^{k(l_{0}+\nu_{0})}(L)\; .
\eeqa
\end{mathletters}
The corresponding finite-size scaling partition function is\footnote{Partition
functions connected to the toroidal XXZ chain (\ref{tham}) have superscript
in addition to subscript, in contrast to those connected to the our $SU(2)_{q}$
invariant chain (\ref{qham}).}
\beqab
{\cal F}_{Q}^{k}(z,\bar{z})&=&\lim_{L\rightarrow\infty}{\cal F}_{Q}^{k}(z,
\bar{z},L) \\ &=&\lim_{L\rightarrow\infty}\sum_{j=1}^{L \choose Q+L/2}
z^{\frac{1}{2}(\bar{F}_{Q;j}^{k}+\bar{P}_{Q;j}^{k})}\,\bar{z}^{
\frac{1}{2}(\bar{F}_{Q;j}^{k}-\bar{P}_{Q;j}^{k})}\;\; . 
\eeqab
According to \cite{grimm,pallua}, one subset of
$c<1$ models compatible with (\ref{mvp}) can be projected by imposing:
\beq \label{lnm}
l_{0}+\nu_{0}=1-\frac{1}{4h}=\frac{1}{m+1}\; .
\eeq
If we define
\beq \label{dpfl}
{\cal D}_{Q}^{k}(z,\bar{z},L)\equiv {\cal F}_{Q}^{k}(z,\bar{z},L)
-{\cal F}_{k}^{Q}(z,\bar{z},L)
\eeq
we obtain \cite{grimm}
\beqa
{\cal D}_{Q}^{k}(z,\bar{z})&\equiv& \lim_{L\rightarrow\infty}
{\cal D}_{Q}^{k}(z,\bar{z},L) \nn \\ &=&\sum_{r=1}^{\infty}
\chi_{r,k-Q}(z)\,\chi_{r,k+Q}(\bar{z}) \; . \label{dpf}
\eeqa
Here $\chi_{r,s}$ are character functions of irreducible representations of
the Virasoro algebra with highest weights $\Delta_{r,s}$ given by
\beq \label{sdm}
\Delta_{r,s}=\frac{[(m+1)r-ms]^{2}-1}{4m(m+1)}\; .
\eeq
We should emphasize here that $k$ gives boundary condition, according to
(\ref{pl}), (\ref{fsg}) and (\ref{lnm}), to be
\beq \label{pkm}
\phi=2\pi l=2\pi k(l_{0}+\nu_{0})=2\pi\frac{k}{m+1}
\eeq

\subsection{Quantum chain and CFT}

Now we are going to make a connection with quantum chain. To do that, first
we must have same Bethe equations. Comparing (\ref{qba}) and (\ref{tba}), and
using (\ref{mvp}) we see that
\beqa
\phi&=&2\varphi(Q+1)\!\!\pmod{2\pi}=2\pi\frac{m(Q+1)}{m+1}\!\!\pmod{2\pi} \nn
\\ &=&-2\pi\frac{Q+1}{m+1}\!\!\pmod{2\pi} \label{pvpq}
\eeqa
Comparing now (\ref{pvpq}) and (\ref{pkm}) it follows that we should take
\beq \label{kq}
k=-(Q+1)
\eeq
If we define
\bdm
H_{Q}^{k}(L)=H(q,\phi,L){\cal P}_{Q}(L)
\edm
where $H(q,\phi,L)$ is toroidal XXZ Hamiltonian (\ref{tham}) (with $\phi$
given by (\ref{pkm})) and ${\cal P}_{Q}$ is projection operator on
sector $S^{z}=Q$, from \cite{pasq} and Appendix of \cite{grimm} follows:
\begin{mathletters} \label{hspl}
\beqa
S^{+}H_{Q}^{-(Q+1)}&=&H_{Q+1}^{-Q}S^{+} \\
S^{+}T_{Q}^{-(Q+1)}&=&\pm T_{Q+1}^{-Q}S^{+}
\eeqa
\end{mathletters}
So, because when we make difference in 
\beqab
{\cal D}_{Q}^{-(Q+1)}&=&{\cal F}_{Q}^{-(Q+1)}-{\cal F}_{-(Q+1)}^{Q}\\
                     &=&{\cal F}_{Q}^{-(Q+1)}-{\cal F}_{Q+1}^{-Q}
\eeqab
(we have used charge simmetry generated by $C=\prod\sigma_{j}^{x}$ which
changes signs of $S_{z}$ and $\phi$, see (2.12) in \cite{grimm})
we make difference between spectra of $H_{Q}^{-(Q+1)}$ and $H_{Q+1}^{-Q}$, 
from (\ref{hspl}) follows that we obtain only highest weight states of
$SU(2)_{q}$. 
And these are exactly solutions of Bethe equations which are the same for both 
chains. Thus the quantum chain partition function is
\beqab
{\cal D}_{Q}(z,\bar{z})&=&\lim_{L\rightarrow\infty}{\cal D}_{Q}(z,
\bar{z},L)\\ &=& {\cal D}_{Q}^{-(Q+1)}(z,\bar{z}) \\ &=&
\sum_{r=1}^{\infty}\chi_{r,-(2Q+1)}(z)\,\chi_{r,-1}(\bar{z}) 
\eeqab
Using the formal relation $\chi_{r,-s}=-\chi_{r,s}$ we can write the final
result
\beq
{\cal D}_{Q}(z,\bar{z})=\sum_{r=1}^{\infty}\chi_{r,2Q+1}(z)\,
\chi_{r,1}(\bar{z})
\eeq
In our case the number $m$ appearing in Virasoro characters (see (\ref{sdm})) 
is connected to the $q=\exp (i\varphi)$ parameter with relation (\ref{mvp}).

\end{document}